\documentclass[12pt,english]{article}
\usepackage[T1]{fontenc}
\usepackage[latin1]{inputenc}
\usepackage{babel}
\usepackage{graphics}
\usepackage{floatflt}

\makeatletter

\providecommand{\LyX}{L\kern-.1667em\lower.25em\hbox{Y}\kern-.125emX\@}
\let\SF@@footnote\footnote
\def\footnote{\ifx\protect\@typeset@protect
    \expandafter\SF@@footnote
  \else
    \expandafter\SF@gobble@opt
  \fi
}
\expandafter\def\csname SF@gobble@opt \endcsname{\@ifnextchar[
  \SF@gobble@twobracket
  \@gobble
}
\edef\SF@gobble@opt{\noexpand\protect
  \expandafter\noexpand\csname SF@gobble@opt \endcsname}
\def\SF@gobble@twobracket[#1]#2{}

\usepackage{graphicx}
\usepackage{here}

\makeatother
\begin{document}

\title{POSSIBLE MEASUREMENTS OF GPDs AT COMPASS%
\thanks{Talk given at the workshop {}``Future physics at COMPASS'', Cern
(2002)
}}

\author{N. d'Hose\( ^{1} \), E. Burtin\( ^{1} \), P.A.M. Guichon\( ^{1} \),
J. Marroncle\( ^{1} \), \\
M. Moinester\( ^{2} \), J. Pochodzalla\( ^{3} \), A. Sandacz\( ^{4} \)}

\maketitle
\begin{abstract}
This paper presents the reactions which can be performed at COMPASS
to study the Generalized Parton Distributions (GPDs). The high energy
muon beam at CERN allows to measure Hard Exclusive Meson Production
or Deeply Virtual Compton Scattering (DVCS) in the Bjorken regime
in a large range of \( Q^{2} \) and \( x_{Bj} \) (1.5 \( \leq Q^{2}\leq  \)
7.5 GeV\( ^{2} \) and 0.03 \( \leq x_{Bj}\leq  \) 0.25). Exploratory
measurements dedicated to \( \rho ^{0} \) or \( \pi ^{0} \) production
can be investigated with the present setup. DVCS measurement require
an upgrade of the COMPASS setup.
\end{abstract}

\section{Goal of an experiment with the high energy muon beam}

The Generalized Parton Distributions (GPDs) provide a unified description
of the nucleon. As explained by M. Diehl~\cite{markus}, they interpolate
between the parton distributions and the hadronic form factors. Experimentally
the GPDs can be accessed in exclusive measurements such as Hard Exclusive
Meson (\( \rho ,\pi ... \)) Production (HEMP) and Deeply Virtual
Compton Scattering (DVCS). The latter reaction is the simplest from
the theoretical point of view but also the most difficult experimentally
because one has to select perfectly the final state (one lepton, one
proton and one photon) among all the possible reactions. In pratice
we can by now start the investigation of Meson Production and we foresee
an upgrade of the COMPASS setup for DVCS measurement.

\begin{floatingfigure}{0.5\textwidth}
{\centering \resizebox*{0.4\textwidth}{!}{\includegraphics{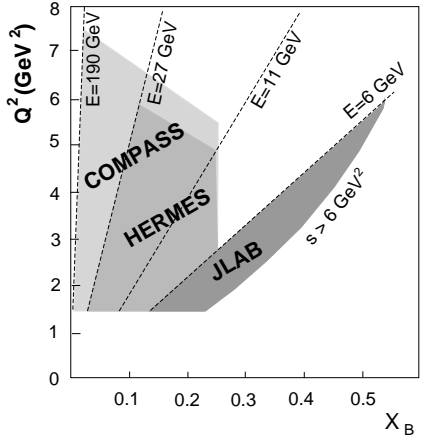}} \par}

\caption{Kinematical coverage for various planned or proposed experiments.
The limit \protect\( s\geq 6\protect \) GeV\protect\( ^{2}\protect \)
assures to be above the resonance domain, and \protect\( Q^{2}>1.5\protect \)
GeV\protect\( ^{2}\protect \) allows to reach the Deep Inelastic
regime.}

\label{kindam}
\end{floatingfigure}
Experiments have already been undertaken at very high energy with
the HERA collider to study mainly the gluon GPDs at very small \( x_{Bj} \)
(\( \leq 10^{-2} \)). Larger values of \( x_{Bj} \) have been investigated
in fixed target experiments at JLab (6 GeV, with plans for an upgrade
at 11 GeV) and HERMES (at 27 GeV). The experimental program using
COMPASS at CERN (at 100 and/or 190 GeV) will enlarge the kinematical
domain to a large range of \( Q^{2} \) and \( x_{Bj} \) (1.5 \( \leq Q^{2}\leq  \)
7 GeV\( ^{2} \) and 0.03 \( \leq x_{Bj}\leq  \) 0.25) (see Fig.
\ref{kindam}). A large range in \( Q^{2} \) is required to control
the factorisation in a hard, pertubatively calculable amplitude and
a soft amplitude which is parametrized by the generalized parton distributions
\( H,E,\tilde{H},\tilde{E} \). The GPDs depend on three kinematical
variables: \( x \) and \( \xi  \) parameterize the longitudinal
momentum fractions of the partons, while \( t \) relates to the transverse
momentum transfer.

\begin{figure}
{\centering \resizebox*{0.5\textwidth}{!}{\includegraphics{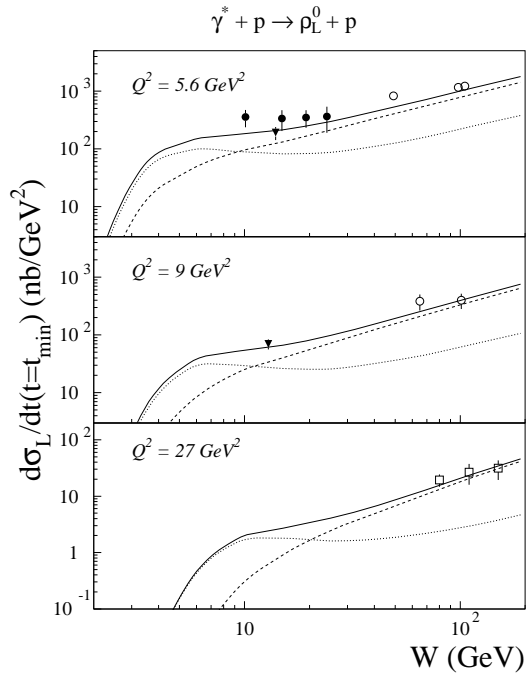}} \par}

\caption{Longitudinal forward differential cross section for \protect\( \rho ^{0}_{L}\protect \)
production (Fig. from~\cite{marcvdh}). Predictions reproduce quark
contributions (dotted lines), gluon contributions (dashed lines) and
the sum of both (full lines). The data are from NMC (triangles)~\cite{NMC},
E665 (solid circles)~\cite{E665}, ZEUS 93 (open circles)~\cite{ZEUS 93}
and ZEUS 95 (open squares)~\cite{ZEUS 95}. }

\label{ rhotot }
\end{figure}

Since the theoretical proof of factorization assumes that the transfer
\( t \) is finite (that is \( t/Q^{2}\rightarrow 0 \)) ~\cite{collins,markus},
we consider in the following \( |t| \) smaller than 1 GeV\( ^{2} \).
Another condition of factorization concerns the helicity of the virtual
photon. In case of Hard Exclusive Meson Production it is mandatory
to impose that the virtual photon be longitudinal in order to select
the perturbative gluon exchange. Experimentally we should consider
Rosenbluth separation for \( \pi ^{0} \) production, while for \( \rho ^{0} \)
production we can select longitudinal \( \rho ^{0} \)s through the
angular distribution of the decay products and assume the s-channel
helecity conservation. Hard Exclusive Meson Productions seem more
complex to analyze as they contain non perturbative information on
both the target and the produced meson. Nevertheless they offer the
possibility to disentangle different GPDs (vector meson production
depends on \( H \) and \( E \) only; pseudo-scalar production depends
on \( \tilde{H} \) and \( \tilde{E} \) only) and to separate contributions
from different flavors. Forward differential longitudinal \( \rho ^{0}_{L} \)
electroproduction cross section measurements which provide the largest
counting rates, have already been undertaken and are presented in
Fig. \ref{ rhotot } as a function of c.m. energy W for three values
of \( Q^{2} \) (5.6, 9 and 27 GeV\( ^{2} \)). The theoretical curve
is an incoherent sum of the quark and gluon contributions~\cite{marcvdh}.
No measurement has been done at \( x_{Bj} \) larger than 0.05. Thanks
to the expertise of the NMC collaboration for these absolute measurements
we will explore a larger domain in \( x_{Bj} \), \( Q^{2} \) and
\( t \) with the muon beam available at CERN.

\begin{figure}
{\centering \resizebox*{0.7\textwidth}{!}{\includegraphics{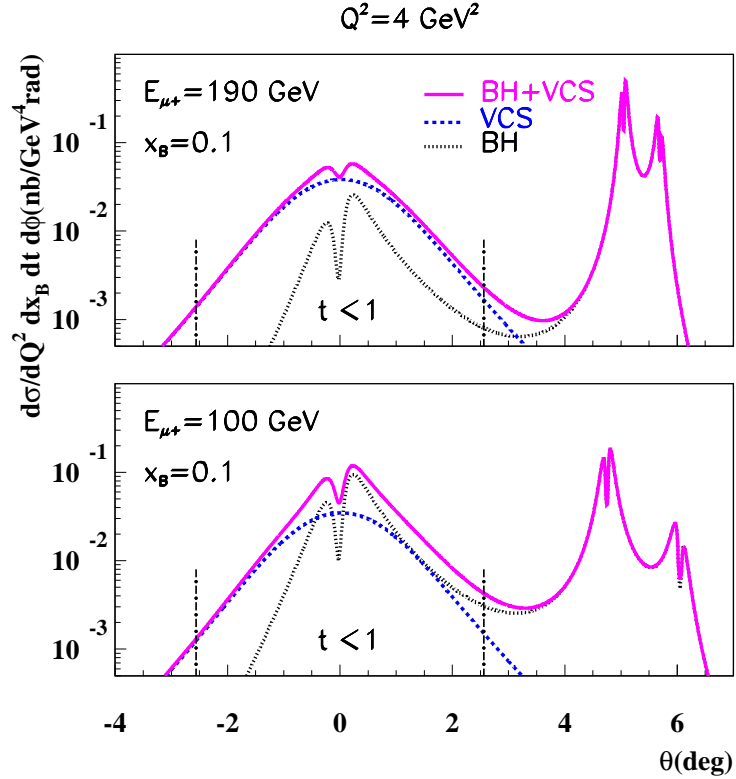}} \par}

\caption{\label{cross-section}Cross sections for the photon leptoproduction
\protect\( \mu p\rightarrow \mu p\gamma \protect \) as a function
of the outgoing real photon angle (relative to the virtual photon
direction). Comparison between BH (dotted lines), DVCS (dashed lines)
and the total cross sections (full lines) for 2 energies of the muon
beam available at CERN: 190 and 100 GeV. The interesting domain is
limited by a transfer \protect\( |t|\protect \) smaller than \protect\( 1{\rm GeV}^{2}\protect \)
i.e. \protect\( \theta \protect \) investigating a small region around
0 degree.}
\end{figure}

Deeply virtual Compton scattering is accessed by photon lepto-production:
\( lp\rightarrow l'p'\gamma  \). In this reaction, the final photon
can be emitted either by the leptons (Bethe-Heitler process) or by
the proton (genuine DVCS process). If the lepton energy is large enough
(see Fig. \ref{cross-section} with \( E_{\mu } \) = 190 GeV, \( Q^{2} \)=
4 GeV\( ^{2} \), \( x_{Bj} \)=0.1), the DVCS contribution dominates
over the BH contribution so that the cross section is essentially
the square of the DVCS amplitude which, at leading order, has the
form: \[
T^{DVCS}\sim \int _{-1}^{+1}\frac{H(x,\xi ,t)}{x-\xi +i\epsilon }dx...\sim \mathcal{P}\int _{-1}^{+1}\frac{H(x,\xi ,t)}{x-\xi }dx...-i\pi H(\xi ,\xi ,t)....\]
 (where \( \xi \sim x_{Bj}/2 \) and \( t \) are fixed by the experiment).
At smaller lepton energy (see Fig. \ref{cross-section} with \( E_{\mu } \)
= 100 GeV and same values of \( Q^{2} \) and \( x_{Bj} \) as above),
the interference between BH and DVCS becomes large and offers a unique
opportunity to study Compton scattering amplitude including its phase.
A careful analysis of the dependence of the cross section on the azimuthal
angle \( \phi  \) between the leptonic and hadronic planes and on
\( Q^{2} \) allows one to disentangle higher-twist effects and to
select the real or imaginary parts of the DVCS amplitude (see the
details in the previous text of M. Diehl~\cite{markus}). If a longitudinally
polarized lepton beam and an unpolarized target are used, the angular
analysis and the \( Q^{2} \) dependence of the cross section difference
\( \sigma (e^{\uparrow })-\sigma (e^{\downarrow }) \) allow one to
select the imaginary part of the DVCS amplitude and thus the GPDs
at the specific values \( x=\xi  \). This study is being investigated
at HERMES~\cite{HERMES} and JLab~\cite{JLab}. If two muon beams
of opposite charge and polarization are used, the angular analysis
and the \( Q^{2} \) dependence of the sum of cross sections \( \sigma (\mu ^{+\downarrow })+\sigma (\mu ^{-\uparrow }) \)
allow also one to select the imaginary part of the DVCS amplitude.
Moreover the same method applied to the difference of cross sections
\( \sigma (\mu ^{+\downarrow })-\sigma (\mu ^{-\uparrow }) \) allows
one to select the real part of the DVCS amplitude which, for a given
\( \xi  \), is sensitive to the complete dependence on \( x \) of
the GPDs. The deconvolution (over \( x \)) of this formula to extract
the GPDs is not yet clearly solved, but comparison to model predictions
can easily be made. It is clear that the muon beam of high energy
at CERN can offer many possibilities in order to investigate the many-faceted
problem of the GPDs knowledge.

\begin{figure}
{\centering \resizebox*{0.7\textwidth}{!}{\includegraphics{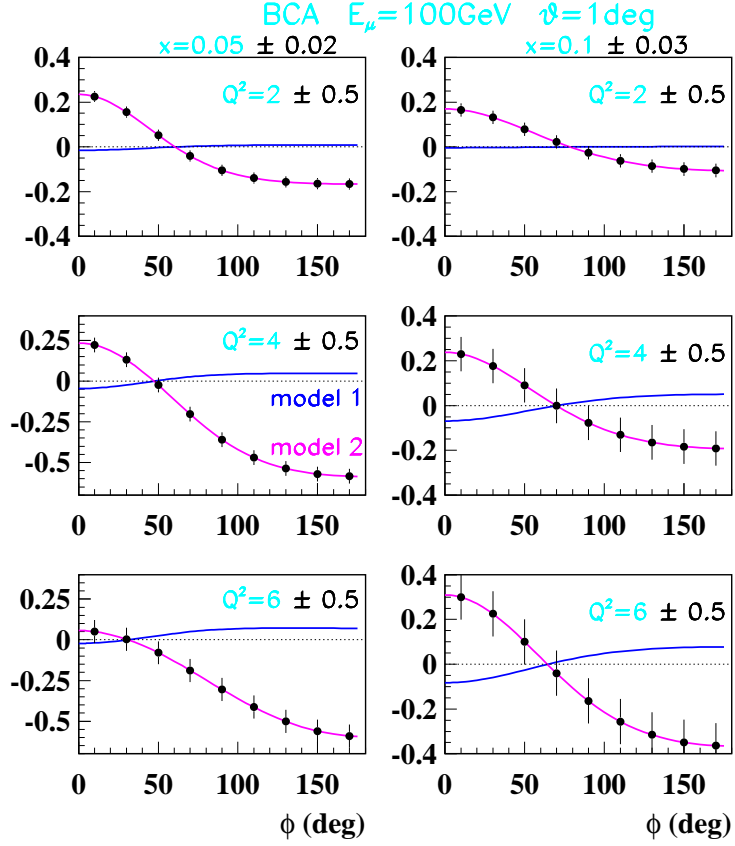}} \par}

\caption{\label{azymuth}Azimuthal distribution of the beam charge asymmetry
measured at COMPASS at \protect\( E_{\mu }\protect \)= 100 GeV and
\protect\( |t|\leq 0.6\protect \) GeV\protect\( ^{2}\protect \)
for 2 domains of \protect\( x_{Bj}\protect \) (\protect\( x_{Bj}=0.05\pm 0.02\protect \)
and \protect\( x_{Bj}=0.10\pm 0.03\protect \)) and 3 domains of \protect\( Q^{2}\protect \)
(\protect\( Q^{2}=2\pm 0.5\protect \) GeV\protect\( ^{2}\protect \),
\protect\( Q^{2}=4\pm 0.5\protect \) GeV\protect\( ^{2}\protect \)
and \protect\( Q^{2}=6\pm 0.5\protect \) GeV\protect\( ^{2}\protect \))
obtained in 6 months of data taking with a global efficiency of 25\%
and with \protect\( 2\cdot 10^{8}\protect \) \protect\( \mu \protect \)
per SPS spill (\protect\( P_{\mu ^{+}}=-0.8\protect \) and \protect\( P_{\mu ^{-}}=+0.8\protect \)) }
\end{figure}

Figure \ref{azymuth} shows the azimuthal distribution of the charge
asymmetry which can be measured at COMPASS and the strong sensitivity
to two different models~\cite{mosse}. The first one is based on
a simple parametrization of the GPDs: \( H^{f}(x,\xi ,t)=H^{f}(x,\xi ,0)F^{f}_{1}(t)/2 \)
where \( F^{f}_{1}(t) \) represents the elastic Dirac form factor
for the quark flavor \( f \) in the nucleon. The second one~\cite{burkardt, pire, diehl,belistsky+mueller}
relies on the fact that the GPDs measure the contribution of quarks
with longitudinal momentum fraction \( x \) to the corresponding
form factor as is suggested by the sum rule: \[
\int ^{+1}_{-1}H^{f}(x,\xi ,t)dx=F^{f}_{1}(t).\]
 As one can associate the Fourier transform of form factors with charge
distributions in position space, one can expect that the GPDs contain
information about the distribution of partons in transverse position
space. In fact it has been demonstrated that, when \( t \) is purely
transverse which amounts to \( \xi =0 \), then \( H(x,0,t) \) is
the Fourier transform of the probability density to find a quark with
momentum fraction \( x \) at a given distance from the center of
momentum in the transverse plane. Qualitatively one expects that quarks
with a large \( x \) come essentially from the small valence {}``core\char`\"{}
of the nucleon, while the small \( x \) region should receive contributions
from the much wider meson {}``cloud\char`\"{}. Therefore one expects
a gradual increase of the \( t \)-dependence of \( H(x,0,t) \) as
one goes from larger to smaller values of \( x \). This suggests
the parametrization: \( H(x,0,t)=q(x)e^{t<b_{\perp }^{2}>}=q(x)/x^{\alpha t} \)
where \( <b_{\perp }^{2}>=\alpha \cdot ln1/x \) represents the increase
of the nucleon transverse size with energy. The domain of small \( x_{Bj} \)
reached at COMPASS is related to the observation of see quarks or
meson {}``cloud\char`\"{} or also gluons and it provides a large
sensitivity to this three-dimensional picture of partons inside a
hadron.

\section{General requirements for COMPASS}

The highest luminosity reachable at COMPASS is required to investigate
these exclusive measurements. The experiment will use 100-190 GeV/c
muons from the M2 beam line. Limits on radio-protection in the experimental
hall imply that the maximum flux of muon to be expected is of \( 2\cdot 10^{8} \)
muons per SPS spill (5.2s spill duration, repetition each 16.8s).
Under these circumstances, we can reach a luminosity of \( \mathcal{L}=5\cdot 10^{32} \)
cm\( ^{-2} \)s\( ^{-1} \) with the present polarized \( ^{6} \)LiD
or NH\( _{3} \) target of 1.2 meter long, and only \( \mathcal{L}=1.3\cdot 10^{32} \)
cm\( ^{-2} \)s\( ^{-1} \) with a future liquid hydrogen target of
2.5 meter long.

In order to get useful cross sections with positive and negative muon
beams, it is necessary to perform a precise absolute luminosity measurement.
This has already been achieved by the NMC Collaboration within a 1\%
accuracy~\cite{abslumiNMC}. The integrated muon flux was measured
continuously by two methods: either by sampling the beam with a random
trigger (provided by the \( \alpha  \) emitter Am\( ^{241} \)) or
by sampling the counts recorded in 2 scintillators hodoscope planes
used to determine incident beam tracks. The beam tracks were recorded
off-line, in the same way as the scattered muon tracks to determine
exactly the integrated usable muon flux.

Moreover \( \mu ^{+} \) and \( \mu ^{-} \) beams of 100 GeV energy,
with the same and as large as possible intensity as well as exactly
opposite polarization (to a few \%) are required. The muons are provided
by pion and kaon decay and are naturally polarized. The pions and
kaons come from the collision of the SPS 400 GeV proton beam on a
Be primary target. A solution was proposed by Lau Gatignon~\cite{lau}.
It consists in: 1) selecting 110 GeV pion beams from the collision
and 100 GeV muon beams after the decay section in order to maximize
the muon flux; 2) keeping constant the collimator settings which define
the pion and muon momentum spreads (both the collimator settings in
the hadron decay section and the scrapper settings in the muon cleaning
section) in order to fix the \( \mu ^{+} \) and \( \mu ^{-} \) polarizations
at exactly the opposite value (\( P_{\mu ^{+}}=-0.8 \) and \( P_{\mu ^{-}}=+0.8 \));
3) fixing \( N_{\mu ^{-}} \) to \( 2\cdot 10^{8} \) \( \mu  \)
per SPS spill with the longest 500mm Be primary target; 4) using a
shorter target to find \( N_{\mu ^{+}} \) close to \( 2\cdot 10^{8} \)
\( \mu  \) per SPS spill.

This paragraph presents the experimental procedure to select the exclusive
HEMP or DVCS channel and the difference equipments that are required.
They are mostly part of the existing high resolution COMPASS spectrometer:
muon detection which insures a good resolution in \( x_{Bj} \) and
\( Q^{2} \), meson detection and identification in RICH or photon
detection in calorimeters of good energy and position resolutions
to allow two photons separation. The COMPASS spectrometer intercepts
only forward outgoing particles (until 10 degrees) and the photon
or meson detection limits the experiment to small \( x_{Bj} \) values
(\( x_{Bj}\leq 0.15 \)). At these high energies the complete final
state, including the low energy recoiling proton, needs to be detected
because missing mass techniques are not efficient due to the experimental
resolutions (the resolution in missing mass which is required is \( (m_{p}+m_{\pi })^{2}-m_{p}^{2}=0.25 \)
GeV\( ^{2} \) and the experimental resolution which can be achieved
is larger than 1 GeV\( ^{2} \)). Consequently the high resolution
COMPASS spectrometer needs to be completed by a recoil detector to
measure precisely the proton momentum and exclude other reactions
under high luminosity conditions. In the next section we will try
to by-pass the necessity of a recoil detector to investigate the cleanest
channel: \( \mu p\rightarrow \mu p\rho ^{0} \) where \( \rho ^{0} \)s
are identified through their decay in two charged pions accurately
measured in the forward COMPASS spectrometer.

\section{A pragmatic solution with the present setup}

With the present COMPASS setup we can undertake Hard Exclusive \( \rho ^{0} \)
Production (with the largest cross section) as we can benefit of the
good expertise for this measurement in the previous NMC and SMC experiments
and so take the advantage to produce data in this field as soon as
possible. At the same time we can also investigate Hard Exclusive
\( \pi ^{0} \) Production to see the limit of this setup.

In this context the constraints and limits of these experiments are
the following ones:\\
 1) With the present polarized \( ^{6} \)LiD or NH\( _{3} \) target
the production occurs on quasi-free nucleons in the nucleus or in
the coherent scattering on the nucleus.\\
 2) Luminosity determination is needed, but can be realized indirectly
by measuring at the same time unpolarized Deep Inelastic Scattering
and using known or calculated structure functions \( F_{2} \) and
\( R \) (the {}``EMC'' nuclear effects have to be taken into account).\\
 3) The selection of longitudinal \( \rho ^{0} \) will be made by
the angular distribution of the decay product. No Rosenbluth separation
is envisageable for \( \pi ^{0} \) production.\\
 4) The absence of a recoil detector prevents the complete exclusivity
of the channel.

\begin{figure}
{\centering \resizebox*{0.7\textwidth}{!}{\includegraphics{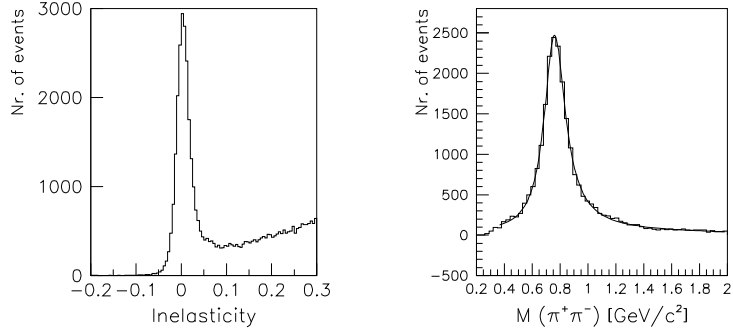}} \par}

\caption{\label{inelastic}The SMC results~\cite{tripet} for \protect\( \mu N\rightarrow \mu \rho ^{0}N\protect \).
(left) Inelasticity distribution after selections; (right) mass spectrum
after selection including the cut \protect\( -0.05\leq I\leq 0.05\protect \).
The full line represents a prediction according the S\"{o}ding model. }
\end{figure}

Precise simulations of exclusive \( \rho ^{0} \) and \( \pi ^{0} \)
production have been performed~\cite{pochodzalla,sandacz} and were
already presented at the COMPASS meeting in Munich in 2000. The selection
of exclusive events can be summarized as follows:\\
 1) Deep inelastic events are selected by cuts on variables depending
on the scattered muon kinematics:\\
 \hspace*{3cm} \( 2\leq Q^{2}\leq Q^{2}_{max} \) and \( 35(20)\leq \nu \leq 170(90) \)
GeV for \( \rho ^{0} \)\\
 \hspace*{3cm} \( 1\leq Q^{2}\leq Q^{2}_{max} \) and \( 15(10)\leq \nu \leq 170(85) \)
GeV for \( \pi ^{0} \)\\
 The values outside(inside) the brackets correspond to the beam energy
of 190(100) GeV.\\
 2) \( \rho ^{0} \) and \( \pi ^{0} \) are identified through decays:
\( \rho ^{0}\rightarrow \pi ^{+}\pi ^{-} \) and \( \pi ^{0}\rightarrow \gamma \gamma  \).
Only two hadrons of opposite charge associated with the vertex defined
by the incident and scattered muons are required for \( \rho ^{0} \)
production and only two photons with the incident and scattered muons
are demanded for \( \pi ^{0} \) production. For \( \rho ^{0} \),
hadrons have to be identified as pions. It is then required that each
pion decay is emitted in the laboratory at an angle smaller than 180
mrad (the acceptance limit of the forward spectrometer) and its momentum
is above 2 GeV. For \( \pi ^{0} \) it is demanded that each decay
photon has energy above 2 GeV and enters either electromagnetic calorimeter
ECAL1 or ECAL2. In addition the separation of two photons at the entrance
of a calorimeter should be larger than 4 cm.\\
 3) To isolate at best the exclusive \( \rho ^{0} \) events, a cut
on the inelasticity I~%
\footnote{\( I=\frac{M_{X}^{2}-M_{p}^{2}}{W^{2}} \) where \( W^{2}=(p+q)^{2} \)
is the total energy squared in the virtual photon-proton system and
\( M_{X^{2}}=(p+q-v)^{2} \) is the missing mass squared of the undetected
recoiling system (\( p \), \( q \), \( v \) are the four-momenta
of the target proton, virtual photon and meson respectively.) 
} is used. In Fig. \ref{inelastic} the inelasticity distribution is
shown for the SMC \( \rho ^{0} \) sample~\cite{tripet} for the
events with the invariant mass in the central part of the \( \rho ^{0} \)
invariant mass peak. For the inelasticity distribution the peak at
I=0 is the signal of exclusive \( \rho ^{0} \) production. Non-exclusive
events, where in addition to detected fast hadrons, slow undetected
hadrons are produced, appear at \( I\geq 0 \). However, due to the
finite resolution, they are not resolved from the exclusive \( \rho ^{0} \)
peak. For the cut \( -0.05\leq I\leq 0.05 \) defining the exclusive
sample the amount of the residual non-exclusive background for the
SMC experiment was up to about 10\% at large \( Q^{2} \).\\
 4) Finally for the \( \rho ^{0} \) channel a cut on the invariant
mass of the two pions can be applied in order to reduce the non-resonant
background. For the SMC sample the invariant mass distribution after
selections, including the cut on inelasticity, is shown in Fig. \ref{inelastic}.
Although the shape of the mass spectrum varies with \( Q^{2} \),
a mass cut, i.e. \( 0.62\leq m_{\pi ^{+}\pi ^{-}}\leq 0.92 \) GeV\( ^{2} \),
allows a selection of \( \rho ^{0} \) events with the relatively
low amount of non-resonant background.\\
 It is clear that the good resolution on charged particles associated
to \( \rho ^{0} \) decay allows the criteria 3 and 4 and thus provides
a good signature of the exclusive \( \rho ^{0} \) channel. The selection
of the \( \pi ^{0} \) channel depends strongly on the quality of
the electromagnetic calorimeters, but constraints cannot be so nicely
determined.

The simulation uses event generators based on a traditional parametrization
on NMC data for \( \rho ^{0} \) production and on 2 models of GPDs
for \( \pi ^{0} \) production. Secondary interactions of the decay
charged pions, absorption of the decay photons in the target, kinematical
smearing based on the experimental resolution, trigger acceptance,
acceptance for pions and photons and track reconstruction efficiency
are considered. A global \( \rho ^{0} \) selection efficiency which
takes into account secondary interactions, three tracks efficiency,
two pion acceptance, a cut on \( I \) and \( M_{\pi ^{+}\pi ^{-}} \)
and muon trigger acceptance is evaluated to 0.21(0.04) with the medium
\( Q^{2} \) trigger (which consists of the {}``middle trigger\char`\"{}
and the {}``ladder trigger\char`\"{} of the muon trigger hodoscopes)
and to 0.36(0.30) with the full \( Q^{2} \) range trigger (which
includes, in addition, the newly implemented {}``outer trigger\char`\"{}).
The values outside(inside) the brackets correspond to the beam energy
of 190(100) GeV. A global \( \pi ^{0} \) selection efficiency which
takes into account secondary interaction, two photon acceptance and
muon trigger acceptance is found close to 0.30.

Total cross sections integrated over the \( Q^{2} \) and \( \nu  \)
acceptance and expected counting rates (with the medium \( Q^{2} \)
trigger) for a period of 150 days (1 year) assuming an overall SPS
and COMPASS efficiency of 25\% are presented in table \ref{tab1}.
The two limits for \( \pi ^{0} \) production correspond to the two
models. About two third of the produced \( \rho ^{0} \) are longitudinally
polarized.
\begin{table}
\begin{tabular}{|c|c|c|c|}
\hline 
&
190 GeV&
100GeV&
\\
\hline 
\( \sigma ^{tot}_{\mu N\rightarrow \mu X} \)&
48nb&
38nb&
\\
\hline 
\( \sigma ^{tot}_{\mu N\rightarrow \mu \rho ^{0}N} \)&
286pb&
250pb&
 \( <Q^{2}>\sim 2.9 \)\\
 \( N^{cuts}_{\mu N\rightarrow \mu \rho ^{0}N}/year \)&
97 Kevents&
15Kevents&
 \( <x_{Bj}\sim 0.034> \)\\
\hline
\hline 
\( \sigma ^{tot}_{\mu N\rightarrow \mu \pi ^{0}N} \)&
1.3 to 5.2 pb&
5.8 to 23 pb&
\( <Q^{2}>\sim 1.6 \)\\
 \( N^{cuts}_{\mu N\rightarrow \mu \rho ^{0}N}/year \)&
625 to 2500 events&
1860 to 7440 events&
\( <x_{Bj}\sim 0.040> \) \\
\hline
\end{tabular}

\caption{\label{tab1}}
\end{table}

\begin{figure}

\caption{\label{nstar}Inelasticity distribution for the simulated \protect\( \mu N\rightarrow \mu \rho ^{0}N\protect \)
events at 190 GeV; (bottom) calculated distribution of the inelasticity
for \protect\( \mu N\rightarrow \mu \rho ^{0}N^{*}\protect \) for
a continuum distribution of \protect\( N^{*}\protect \) masses. }

{\centering \resizebox*{0.7\textwidth}{!}{\includegraphics{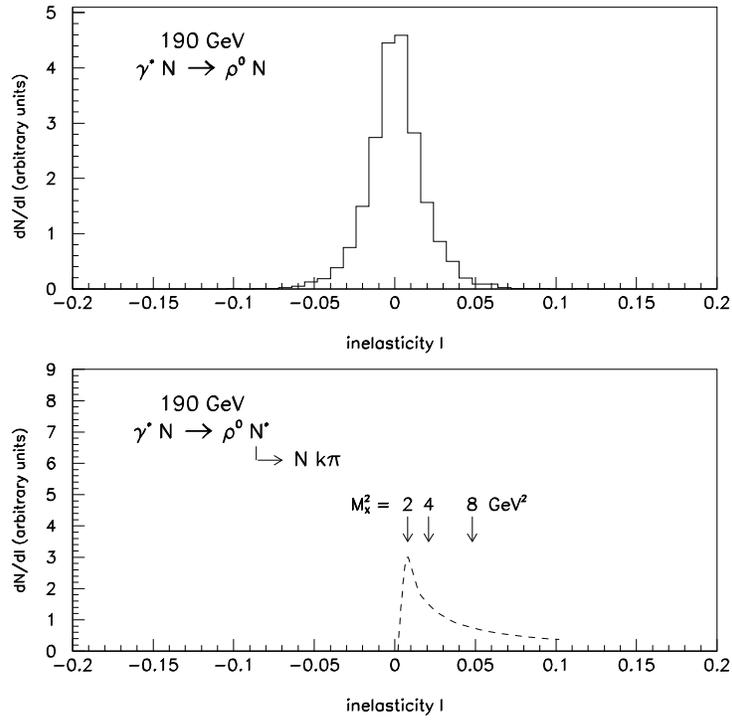}} \par}
\end{figure}

The background to the reaction \( \mu N\rightarrow \mu \rho ^{0}N \)
has been studied. It is due to the events with at least two slow undetected
particles which are outside the acceptance of the spectrometer or
for which tracks are not reconstructed due to inefficiency and which
pass all selections for exclusive \( \rho ^{0} \) events. A first
possible source of this background is \( \rho ^{0} \) production
with diffractive dissociation of the target \( \mu N\rightarrow \mu \rho ^{0}N^{*} \)
with the subsequent decay of the excited state \( N^{*} \) in \( N+k\pi  \).
In the lower part of Fig. \ref{nstar} a schematic drawing of the
inelasticity distribution for these events is presented. Although
the smearing effects are not taken into account it is clear that the
events from diffractive dissociation of the target into the lowest
masses excitations of the nucleon will contribute to the exclusive
\( \rho ^{0} \) sample in the inelasticity cut. An estimation has
been found close to 20\%.\\
 A second source of background will originate from the large number
of inclusive deep inelastic events. Simulation with a sample of 5000
events only using the generators LEPTO and JETSET has given an upper
limit of contamination of also 20\%.

\section{An ideal solution with a completed setup}

As mentioned in the previous section it is clear that only a recoil
detector which allows the low energy recoiling proton detection will
help to select exclusive channels as HEMP or DVCS. The latter reaction
is surely the most delicate because one has to select a final state
with one muon, one photon and one low energy proton among many competing
reactions listed below:\\
 1) Hard Exclusive \( \pi ^{0} \) Production \( \mu p\rightarrow \mu p\pi ^{0} \)
where \( \pi ^{0} \) decays in two photons, for which the photon
with higher energy imitates a DVCS photon, and the photon with smaller
energy is emitted at large angle outside of the acceptance or its
energy is below the photon detection threshold. \\
 2) Diffractive dissociation of the proton \( \mu p\rightarrow \mu \gamma N^{*} \)
with the subsequent decay of the excited state \( N^{*} \) in \( N+k\pi  \).
(The low energy pions are emitted rather isotropically).\\
 3) Inclusive Deep Inelastic Scattering with, in addition to the reconstructed
photon, other particles produced outside the acceptance or for which
tracks are not reconstructed due to inefficiency.\\
 Moreover one has to take into account a background which includes
beam halo tracks with hadronic contamination, beam pile-up, particles
from the secondary interactions and external Bremsstrahlung.

\begin{figure}
{\centering \resizebox*{0.7\textwidth}{!}{\includegraphics{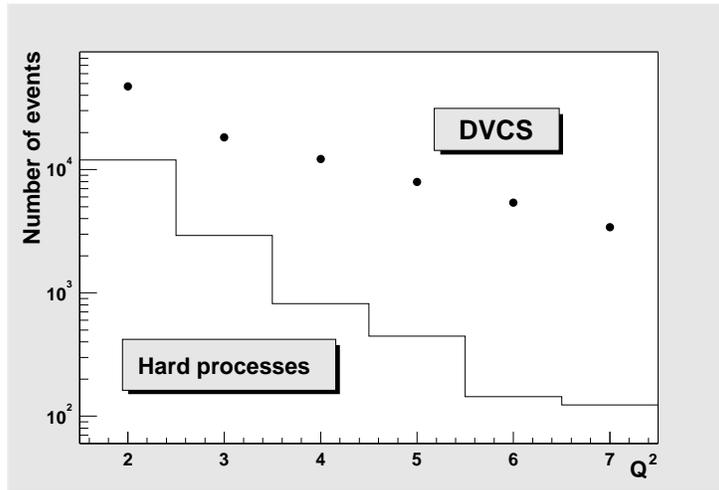}} \par}

\caption{\label{histo}Number of events for DVCS (dots) and DIS (histogram)
processes as a function of \protect\( Q^{2}\protect \) for selection
of events with only one muon, one photon and one recoiling proton
and condition for charged particle detection up to 40 degrees and
for photon detection up to an angle of 24 degrees and above a threshold
of 50 MeV.}
\end{figure}

A simulation has been realized in order to define the proper geometry
of the detector complementing the present COMPASS setup and to analyze
the operational conditions. The goal was to maximize the ratio of
DVCS events over DIS events for a sample of events with one muon and
one photon in the COMPASS spectrometer acceptance plus only one proton
of momentum smaller than 750 MeV/c and angle larger than 40 degrees
(it is the typical kinematics of a DVCS event at small \( t \)).
The simulation relies on the event generator program PYTHIA 6.1~\cite{pythia}
which includes most of the known processes~\cite{friberg} such as
Deeply Inelastic Scattering and Deeply Meson Production. The experimental
parameters such as maximum angle and energy threshold for photon detection
and maximum angle for charged particle detection could then be tuned.
With photon detection extended up to 24 degrees and above an energy
threshold of 50 MeV and with charged particle detection up to 40 degrees,
one observes that the number of DVCS events as estimated with models
is more than an order of magnitude larger than the number of DIS events
over the whole useful \( Q^{2} \) range (see Fig. \ref{histo}). 

The COMPASS setup will be instrumented with two electromagnetic calorimeters
ECAL1 and ECAL2~\cite{compass,calo1}. They are mainly constituted
of lead-glass blocks called GAMS. They are cells of 38.4 \( \times  \)
38.4 \( \times  \) 450 mm\( ^{3} \). Typical characteristics of
such calorimeter are:\\
 - energy resolution: \( \sigma P_{\gamma }/P_{\gamma }=0.055/\sqrt{P_{\gamma }} \)
+ 0.015\\
 - position resolution: \( \sigma _{x}=6.0/\sqrt{P_{\gamma }}+0.5 \)
in mm\\
 - high rate capability: 90\% of signal within 50ns gate with no dead
time\\
 - effective light yield: about 1 photoelectron per MeV; hence low
energy photons of down 20 MeV can be reconstructed.\\
 The separation of the overlapping electromagnetic showers in the
cellular GAMS calorimeter is carefully studied in the Ref.~\cite{calo2}.
The result of the study shows that at 10 GeV one can reach a 100\%
level of the separation efficiency for a minimum distance between
2 photon tracks at the entrance of the calorimeter of D = 4 cm. The
last value is slightly shifted to D = 5 cm at 40 GeV.\\
 This excellent performance of the calorimeters will provide a key
role in the perfect separation between DVCS events and Hard \( \pi ^{0} \)
events.

\begin{figure}
{\centering \resizebox*{0.8\textwidth}{!}{\includegraphics{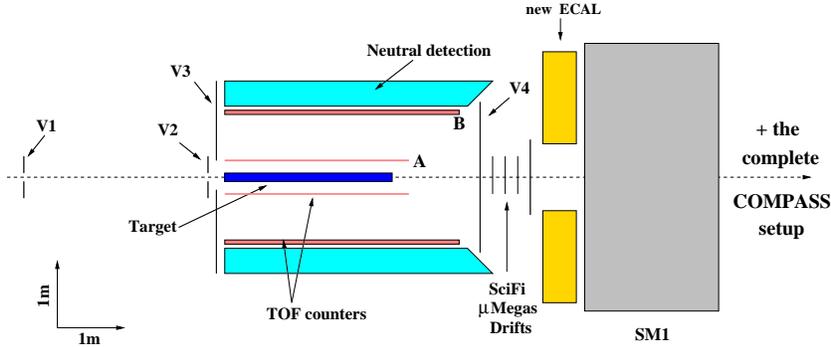}} \par}

\caption{\label{compass}Proposition for a detector complementing the COMPASS
setup. A recoil detector, an extended calorimetry from 10 to 24 degrees,
and a veto (V4) for charged forward particles until 40 degrees have
been added. }
\end{figure}

One possible solution to complement the present COMPASS setup is presented
in Fig. \ref{compass}. It consists of one recoil detector described
below, an extended calorimetry from 10 to 24 degrees, and a veto for
charged forward particles until 40 degrees. This calorimeter has to
work in a crowdy environment and in a magnetic fringe field of SM1
and therefore it has to be studied further.

At the present time our studies have focused on the possibility to
design and successfully operate a dedicated recoil detector. One goal
is to identify and measure the protons momenta between a minimum value
and 750 MeV/c. A solution consists in a large time of flight setup
between a thin segmented cylindrical layer of scintillator counters,
about 3m long and surrounding the 2.5 meter long target, and a thick
layer at about 1m distance from the first layer. The thickness of
the first layer has to be as small as possible in order to detect
protons of minimum momentum. With an hydrogen target of 3cm diameter,
target wall thickness of about 3mm of equivalent scintillator and
a first layer of 4mm, a minimum momentum of 270 MeV/c is reached.
All the counters are read at both sides by photomultiplier counters
to determine time and position with very accurate resolutions (300ps
and 1.8cm). The consequent resolution in momentum varies from 3 to
10\%. The resolution in \( t \) is twice this value, thus it is very
desirable to further study all the parameters which can be improved.
Moreover the exclusion of extra particles have to be studied with
kinematical fits depending on the experimental resolution and/or with
low energy \( \pi ^{0} \) detection. This detector has to work in
a high rate environment. It has to be as large and hermetic as possible
within a reasonable cost. The actual realisation of such performances
is under active investigation.

Counting rates, given in table \ref{tab2}, have been estimated assuming
6 months of data taking (1 year) assuming an overall efficiency of
25\% and considering the present COMPASS setup where the photon detection
is limited to 10 degrees plus a proton detection from 250 to 750 MeV/c
(\( |t|\leq 0.64 \) GeV\( ^{2} \)). This statistics allows for the
\( \phi  \) distribution presented in Fig. \ref{azymuth} for \( Q^{2} \)
= 2, 4 and 6 GeV\( ^{2} \) and \( x_{Bj} \) = 0.05 and 0.10. Studies
devoted to the \( t \) dependence of the cross section can be investigated
but for that it is quite worthwhile to try to improve the \( t \)
resolution.

\begin{table}
\begin{tabular}{|c|c|c|c|}
\hline 
\multicolumn{4}{|c|}{\( E_{\mu } \) = 190 GeV}\\
\hline
&
 \( x_{Bj}=0.05\pm 0.02 \)&
 \( x_{Bj}=0.10\pm 0.03 \)&
 \( x_{Bj}=0.20\pm 0.07 \)\\
\hline
\( Q^{2}=2\pm 0.5 \)&
 10058 events &
 8897 events &
 2000 events \\
 \( Q^{2}=3\pm 0.5 \)&
 3860 events &
 2540 events &
 1300 events \\
 \( Q^{2}=4\pm 0.5 \)&
 2058 events &
 1136 events &
 600 events \\
 \( Q^{2}=5\pm 0.5 \)&
 1472 events &
 677 events &
 520 events \\
 \( Q^{2}=6\pm 0.5 \)&
 875 events &
 459 events &
 357 events \\
 \( Q^{2}=7\pm 0.5 \)&
 642 events &
 299 events &
 242 events \\
\hline
\multicolumn{4}{|c|}{\( E_{\mu } \) = 100 GeV}\\
\hline
&
 \( x_{Bj}=0.05\pm 0.02 \)&
 \( x_{Bj}=0.10\pm 0.03 \)&
 \( x_{Bj}=0.20\pm 0.07 \)\\
\hline
\( Q^{2}=2\pm 0.5 \)&
 13670 events &
 9921 events &
 4300 events \\
 \( Q^{2}=3\pm 0.5 \)&
 5933 events &
 3200 events &
 2000 events \\
 \( Q^{2}=4\pm 0.5 \)&
 4532 events &
 1537 events &
 770 events \\
 \( Q^{2}=5\pm 0.5 \)&
 3000 events &
 995 events &
 600 events \\
 \( Q^{2}=6\pm 0.5 \)&
 1806 events &
 885 events &
 499 events \\
 \( Q^{2}=7\pm 0.5 \)&
 810 events &
 870 events &
 352 events  \\
\hline
\end{tabular}

\caption{\label{tab2}}
\end{table}

\begin{figure}
{\centering \resizebox*{0.9\textwidth}{!}{\includegraphics{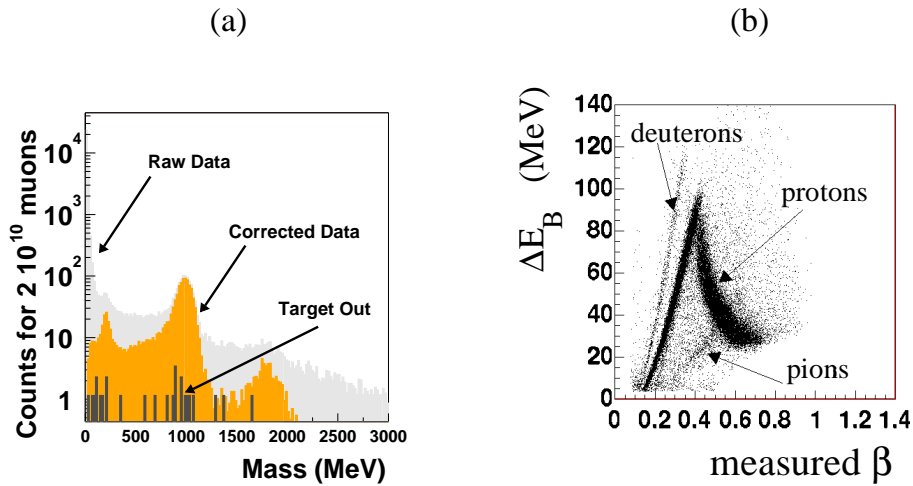}} \par}

\caption{\label{last-figure}(a): Energy lost in the B scintillator as a function
of the measured \protect\( \beta \protect \). (b): Mass distribution
of particles stopped in B. The three peak are pions, protons and deuterons
respectively.}
\end{figure}

We have tested the concept of this detector using the already existing
muon beam and a simplified setup (one sector of scintillators with
reduced length). The muon beam was scattered off a 10~cm long polyethylene
target, mostly equivalent in radiation length to the foreseen long
liquid hydrogen target. We used three scintillators read-out at both
sides, a 4mm~thick close to the target (A), a 5~cm~thick 80~cm
away from the target (B) and an extra scintillator (C) to know if
particles go through B or are stopped in B. The rates observed in
the scintillator close to the target, using the nominal intensity
of \( 2\cdot10 ^{8} \) muons per spill, is of the order of 1~MHz
(mainly due to M\"{o}ller electrons). It demonstrates that the background
environment is acceptable for the time of flight system.\\
 The result of the time of flight operation (see Fig. \ref{last-figure}(a))
shows a clear proton signal. With the knowledge of the \( \beta  \)
velocity and the energy lost in B for stopped particles one can reconstruct
their masses. It is done in Fig. \ref{last-figure}(b) where one can
see pions, protons and deuterons for raw data, corrected data and
the target out contribution which is about two order of magnitude
smaller.

The position resolution obtained on A and B and the time of flight
resolution are better than 2 cm and 300 ps respectively. Extension
to long (3 meter) and thin scintillators have to be studied carefully
and technology has to be improved to achieve still better resolution.
An efficiency study of such a recoil detector is being performed.

\section{CONCLUSIONS}

This study takes advantage of the high energy of the muon beam available
at CERN which provides a large \( Q^{2} \) and \( x_{Bj} \) range
and encourages us for the following roadmap. Hard Exclusive Meson
Production have to be undertaken as soon as possible with the present
setup. A large number of \( \rho ^{0} \) events (a few 10K) can be
produced in 1 year. The \( \rho ^{0} \) channel which decays in \( \pi ^{+}\pi ^{-} \)
is the easiest channel to isolate, the \( \pi ^{0} \) channel is
more difficult but very important to test the calorimetry performances.
A complete experiment with both Hard Exclusive Meson Production with
a large set of mesons and Deeply Virtual Compton Scattering has to
be envisaged in a next step with a completed COMPASS setup. For this
purpose one needs a {}``long\char`\"{} hydrogen target, a recoil
detector and an extension of the calorimetry at larger angles.

COMPASS is the unique place which provides \( \mu ^{+} \) and \( \mu ^{-} \)
of 100 GeV in order to study carefully two scales of observation \( x_{Bj}=0.05\pm 0.02 \)
and \( x_{Bj}=0.10\pm 0.03 \) on a large domain of \( Q^{2} \) from
2 to 7 GeV\( ^{2} \) and to measure the azimuthal distribution of
the Beam Charge Asymmetry which seems very promising to test the geometrical
interpretation of GPDs.

\section*{ACKNOWLEDGMENTS}

We acknowledge useful discussions with Dietrich von Harrach, Fritz
Klein, Alain Magnon and all members of the Saclay COMPASS group.


\begin{thebibliography}{10}
\bibitem{markus}M. Diehl, Introduction to Generalized Parton Distributions, Contribution
to these Proceedings. 
\bibitem{collins}J.C. Collins, L. Frankfurt and M. Strikman, Phys. Rev. \textbf{D56}
(1997) 2982. 
\bibitem{marcvdh}M. Vanderhaeghen, P.A.M. Guichon, M. Guidal, Phys. Rev. \textbf{D
60} (1999) 094017. 
\bibitem{NMC} NMC Collaboration, M. Arneodo \textit{et al.}, Nucl. Phys. \textbf{B429}
(1994) 503. 
\bibitem{E665} E665 Collaboration, M.R. Adams \textit{et al.}, Z. Phys. \textbf{C74}
(1997) 237. 
\bibitem{ZEUS 93} ZEUS Collaboration, M. Derrick \textit{et al.}, Phys. Lett. \textbf{B356}
(1995) 601. 
\bibitem{ZEUS 95} ZEUS Collaboration, J. Breitweg \textit{et al.}, Eur. Phys. J. \textbf{C6}
(1999) 603. 
\bibitem{HERMES} HERMES Collaboration, A. Airapetian \textit{et al.}, Phys. Rev. Lett.
\textbf{87} (2001) 182001 
\bibitem{JLab} CLAS Collaboration, S. Stepanyan \textit{et al.}, Phys. Rev. Lett.
\textbf{87} (2001) 182002 
\bibitem{mosse} L. Moss\'{e}, P.A.M. Guichon, M. Vanderhaeghen, private communication. 
\bibitem{burkardt} M. Burkardt, Phys. Rev. \textbf{D62} (2000) 07503; hep-ph/0207047. 
\bibitem{pire}J.P. Ralston and B. Pire, hep-hp/0110075. 
\bibitem{diehl}M. Diehl, Eur. Phys. J. \textbf{C25} (2002). 
\bibitem{belistsky+mueller}A.V. Belitsky, D. M\"{u}ller, Nucl. Phys. \textbf{A711} (2002) 118;
hep-hp/0206306. 
\bibitem{abslumiNMC} NMC Collaboration, P. Amaudruz \textit{et al.}, Phys. Lett. \textbf{B295}
(1992) 159; R.P. Mount, Nucl. Instrum. Methods \textbf{187} (1981)
401. 
\bibitem{lau} L. Gatignon, private communication. 
\bibitem{pochodzalla} J. Pochodzalla, L. Mankiewicz, M. Moinester, G. Piller, A. Sandacz,
M. Vanderhaeghen, hep-ex/9909534. 
\bibitem{sandacz} A. Sandacz, COMPASS Note 2000-1 (2000). 
\bibitem{tripet} A. Tripet, presented at the 7th International Workshop on DIS and
QCD, Nucl. Phys. \textbf{B}, Proc. Suppl. \textbf{79} (1999). 
\bibitem{pythia} PYTHIA 6.1, User's manual, T. Sj\"{o}strand \textit{et al.}, High
Energy Physics Event Generation with PYTHIA 6.1, Comput. Phys. Commun.
135 (2001) 238; hep-ph/0010017. 
\bibitem{friberg}C. Friberg and T. Sj\"{o}strand, hep-ph/0007314. 
\bibitem{compass} A Proposal for a Common Muon and Proton apparatus for Structure and
Spectroscopy, CERN/SPSLC 96-14 and wwwcompass.cern.ch. 
\bibitem{calo1} V. Poliakov, Presentation of ECAL1 and ECAL2, January 25, 2001. 
\bibitem{calo2}A.A. Lednev, Separation of the overlapping electromagnetic showers
in the cellular GAMs-type calorimeters, Preprint IHEP 93-153 (1993),
Protvino, Russia.\end{thebibliography}
\end{document}